\documentstyle[aps,prb,multicol,epsf]{revtex}
\newcommand{\lineright}{\rule[-1ex]{0.1mm}{1ex}\rule{0.485\linewidth}{0.1mm}}
\newcommand{\lineleft}{\rule{0.485\linewidth}{0.1mm}\rule[0mm]{0.1mm}{1ex}}
\newcommand{\rr}{{\bf r}}
\newcommand{\vs}{{\bf s}}
\newcommand{\vQ}{{\bf Q}}
\newcommand{\vbr}{{\bf R}}
\newcommand{\vk}{{\bf k}}
\newcommand{\vm}{{\bf m}}
\newcommand{\vd}{\mbox{\boldmath $\delta$}}
\newcommand{\djy}{\delta^y_j}

\newcommand{\phij}{\varphi_j}
\newcommand{\phijp}{\varphi_{j'}}
\begin{document}
\draft
\title{SO(5) theory of insulating vortex cores in high-$T_c$ materials}
\author{Brian M\o ller Andersen, Henrik Bruus and Per Hedeg\aa rd}
\address{\O rsted Laboratory, Niels Bohr Institute for APG, \\
Universitetsparken 5, DK-2100 Copenhagen \O\ Denmark}
\date{June 12, 1999}
\maketitle
\begin{abstract}
We study the fermionic states of the antiferromagnetically
ordered vortex cores predicted to exist in the superconducting
phase of the newly proposed SO(5) model of strongly correlated
electrons. Our model calculation gives a natural explanation of the
recent STM measurements on BSCCO, which in surprising contrast to
YBCO revealed completely insulating vortex cores.
\end{abstract}
\begin{multicols}{2}
%\section{Introduction}
The SO(5) model is being developed as a candidate theory for the
high-$T_c$ superconductors. In the seminal paper by Zhang \cite{zhang}
the emphasis was on the phase diagram and the collective (bosonic)
modes of the system. The first experimental support or in
some sense the motivation for the model was the explanation of the 41~meV
magnetic resonance in the superconducting state of YBCO observed by
Mook {\em et al.\/}\cite{mook} Later the energetics of the
normal-superconductivity transition was shown to explain experimental
data on the condensation energy.\cite{condens} More recently the
foundation of the SO(5) model has been clarified by work of Rabello
{\em et al.\/}\cite{rabello},
Henley\cite{henley}, and Eder {\em et al.\/}\cite{eder} These basic
considerations provided the foundation for
studying the Fermi-sector of the SO(5) model, and confirmed the
well known electronic spectrum of the antiferromagnetic (AF) and the
$d$-wave superconducting (dSC) phases. However, the fermionic sector
has not been treated in detail, and the transport properties in the
normal state still remains unresolved.

In this paper we expand on these ideas by studying the fermionic
sector of the SO(5) model in the superconducting phase in the presence
of Abrikosov vortices induced by an external magnetic field. It has
been suggested by Arovas {\em et al.\/}\cite{arovas} that in the SO(5)
model the vortex cores can become insulating, in fact
antiferromagnetic, in stark contrast to the standard normal metal
cores of traditional superconductors. This remarkable
prediction can be tested experimentally by using the improved STM
technique to directly measure the local electronic density of states
in Abrikosov vortices.\cite{hess} The metal cores predicted
by Caroli {\em et al.\/} \cite{caroli} have been seen both in standard
$s$-wave superconductors\cite{hess} (sSC) and in the high-$T_c$ dSC
YBCO.\cite{fischer} The experimental advances naturally led to
intensified theoretical studies of SC vortices. Following initial
calculations on sSC vortex cores,\cite{shore,gygi} the focus soon
turned to dSC cores,\cite{volovik,wang,franz,yasui} and it was
concluded that they are indeed metallic with states very close to the
Fermi energy. Therefore, it was a surprise when Renner {\em et al.\/}
observed that the vortex cores of the high-$T_c$ superconductor BSCCO
was completely devoid of low-lying electronic
excitations.\cite{renner} In this paper we will offer an
explanation of this puzzling experimental observation by solving for the
fermionic sector of the SO(5) model with insulating AF vortex cores.

%\section{Model Hamiltonian}
In our model calculation we consider strongly correlated electrons hopping
on a 2D square lattice with a lattice constant of unit
length. The non-interacting Hamiltonian, $H_0$, is given by an
isotropic tight-binding model:
\begin{equation} \label{H0}
H_0 = -t \sum_{\rr\sigma} \sum_{j=1}^4
c_{\sigma}^{\dagger}(\rr+\vd_j) c_{\sigma}(\rr)
e^{-i\frac{e}{\hbar}\int_{\rr}^{\rr+\delta_j}\!
{\bf A}\cdot d{\bf l}} -\mu \hat{N}
\end{equation}
where $\vd_j = \{\delta_j^x,\delta_j^y\} =
\{\cos[\pi(j-1)/2],\sin[\pi(j-1)/2]\}$ points to the four nearest
neighbors, and where $c_{\sigma}(\rr)$ annihilates an electron with
spin $\sigma$ on site $\rr$. For the interactions the spinor
formalism\cite{rabello} makes it particularly simple to construct a
SO(5) invariant Hamiltonian. In real space it is natural to consider
the spinor
\begin{equation}\label{so5spinor}
  \Psi^{\dagger}(\rr) = \{
  c^{\dagger}_{\uparrow}(\rr),
  c^{\dagger}_{\downarrow}(\rr),
  d_{\uparrow}(\rr),
  d_{\downarrow}(\rr)
  \}.
\end{equation}
The $d_\sigma(\rr)$ operators are associated with
the sites on the opposite sublattice of the one to which $\rr$
belongs:
\begin{equation}\label{ddef}
  d_{\sigma}(\rr) = e^{-i \vQ\cdot\rr} \sum_{\rr}
\varphi(\vbr)
  c_{\sigma}(\rr+\vbr),
\end{equation}
where $\vQ=(\pi,\pi)$, and where $\varphi(\vbr)$ is given by
\begin{equation}\label{phi}
\varphi(\vbr)\!=\!\sum_{\vk}\!e^{i\vk\cdot\vbr} \mbox{sign}(\cos
k_x\!-\!\cos k_y) = \frac{2}{\pi^2}
\frac{1\!-\!e^{i\vQ\cdot\vbr}}{R_x^2\!-\!R_y^2},
\end{equation}
which is only non-zero on the sublattice not including the origin.
The long range nature of $\varphi(\vbr)$ is crucial
for the existence of strict SO(5) symmetry. $\Psi(\rr)$ transforms
like a spinor under SO(5) transformations, {\it i.e.\/} under
rotations in the $ab$ plane generated by the operators $ L_{ab} =
\frac{1}{8} \sum_{\rr}
\Psi^{\dagger}(\rr) \Gamma^{ab} \Psi(\rr)$, $a,b = 1,2,3,4,5$,
where $\Gamma^{ab} \equiv -i \left[ \Gamma^a,\Gamma^b \right]$.
The five $4\times4$ $\Gamma^a$-matrices are given in terms of tensor
products of the standard $2\times2$ Pauli matrices:
$\Gamma^1 = \sigma_y\otimes\sigma_y$,
$\Gamma^x = {\bf I}\otimes\sigma_x$,
$\Gamma^y = \sigma_z\otimes\sigma_y$,
$\Gamma^z = {\bf I}\otimes\sigma_z$, and
$\Gamma^5 = \sigma_x\otimes\sigma_y$. The indices
2, 3, and 4 are writen as $x$, $y$, and $z$ refering to the real
space directions of the AF order parameter. It can be
shown\cite{rabello} that $L_{15}$ corresponds to the charge counting
operator $Q$, that $L_{yz}$, $L_{zx}$, and $L_{xy}$ correspond to the
spin operators $S_x$, $S_y$, and $S_z$, and that $L_{1(x,y,z)}$ are
related to the $\pi_{(x,y,z)}$-operators rotating between the dSC and
AF sectors. As in Ref.~\onlinecite{rabello} we now focus on the vector
interaction, which in the real space representation takes the form
\begin{equation} \label{Hint}
H_{\rm int} = \sum_{a\rr\vs}
V(\rr\!-\!\vs)
\left\{ \Psi^{\dagger}(\rr) \Gamma^{a} \Psi(\rr) \right\}\!
\left\{ \Psi^{\dagger}(\vs) \Gamma^{a} \Psi(\vs) \right\}\!.
\end{equation}
In reality the SO(5) symmetry is broken. However, both the
interpretation of the 41~meV excitation as a pseudo Goldstone mode
relating to a rotation of the dSC phase into the AF phase, as well
as the fact that the coupling strengths in the dSC and AF sectors
are almost identical,\cite{bruus} makes it plausible that the
SO(5) breaking is weak. A natural way to break the SO(5) is simply
to truncate the long range correlations apparent in
Eqs.~(\ref{ddef}) and~(\ref{phi}) for the $d$-operators,
\begin{equation} \label{dtilde}
  d_{\sigma}(\rr) \; \longrightarrow \;
  \tilde{d}_{\sigma}(\rr) =
  \frac{1}{2} e^{-i \vQ\cdot\rr} \sum_{j=1}^4
  \phij c_{\sigma}(\rr+\vd_j),
\end{equation}
where $\phij = (-1)^{\djy}$.
Maintaining only nearest neighbors in the $d$-operator sum constitutes
a simple form relating both to SO(5) symmetry and to Hubbard-like
models for dSC\cite{wang}. Now follow three approximations. First, we
use the truncated $\tilde{d}$-operators instead
of the $d$-operators in the interaction Hamiltonian Eq.~(\ref{Hint}).
Second, we assume a point interaction, $V(\rr-\vs) = -
\frac{1}{8}V \delta(\rr-\vs)$. And third, we utilize
the standard mean-field approximation. These approximations result in
the following SO(5) symmetry broken mean-field interaction
Hamiltonian:
\begin{equation} \label{Hmfint}
  H_{\rm int}^{\rm mf} \!=\! - \!\!\sum_{\rr} \!V\!\left[
  \vm(\rr)\!\cdot\!\langle\vm(\rr)\rangle \!+\! 2 \!\left\{
  \Delta(\rr) \langle\Delta^{\dagger}(\rr)\rangle \!+\! {\rm h.c.}\!
  \right\} \right]\!
\end{equation}
with the dSC and AF order parameters given by
\begin{equation}
\label{Dr}
  \Delta(\rr) = \sum_{j=1}^4 \frac{\phij}{4}  \left\{
  c_{\uparrow}(\rr+\vd_j)   c_{\downarrow}(\rr) -
  c_{\downarrow}(\rr+\vd_j) c_{\uparrow}(\rr) \right\}.
\end{equation}
\begin{eqnarray} \label{mr}  \nonumber
  \vm(\rr) &=& \frac{1}{2} e^{i\vQ\cdot\rr} \left[ \left(
  c^{\dagger}_{\uparrow}(\rr),c^{\dagger}_{\downarrow}(\rr)\right)
  \mbox{\boldmath{$\sigma$}}
  \left( \begin{array}{c} c_{\uparrow}(\rr)\\ c_{\downarrow}(\rr)
  \end{array} \right) \right. \\ && \qquad - \left. \left(
  \tilde{d}^{\dagger}_{\uparrow}(\rr),
  \tilde{d}^{\dagger}_{\downarrow}(\rr)\right)
  \mbox{\boldmath{$\sigma$}}
  \left( \begin{array}{c} \tilde{d}_{\uparrow}(\rr)\\
  \tilde{d}_{\downarrow}(\rr)
  \end{array} \right)
  \right],
\end{eqnarray}
We find that in $H_{\rm int}^{\rm mf}$ of Eq.~(\ref{Hmfint}) the SO(5)
symmetry is broken in such a way that a $d$-wave gap function results
in the pure SC phase, $E_{\vk}^2 = \varepsilon_{\vk}^2 +
[2V|\Delta|(\cos k_x - \cos k_y)]^2$, while a full gap develops in the
pure AF phase, $E_{\vk}^2 = \varepsilon_{\vk}^2 +
[\frac{1}{2}V m(1+(\cos k_x - \cos k_y)^2]^2$.

%-----------------------------------------------------------------------
%\section{The SC-AF interface in SO(5)}
%\label{sec:dSC_AF}
%-----------------------------------------------------------------------
To elucidate the role of the gap in the AF sector we first study
the continuum limit of our model.
The important low-lying excitations in the fermionic sector
are concentrated in the regions near the four $d$-wave gap nodes
$\vQ_{\lambda} = {\scriptstyle \frac{\pi}{2}} (\cos[\pi(\lambda/2 -
1/4)],\sin[\pi(\lambda/2-1/4)])$,
where $\lambda = 1$, 2, 3, 4. We get rid of the rapid variations by
local gauge transformations in each of the four quadrants $\lambda$ in
$\vk$-space:
%\begin{equation} \label{gauge}
$
c_{\sigma}(\rr) = \sum_{\lambda}
e^{i\vQ_\lambda\cdot\rr} \psi_{\lambda,\sigma}(\rr).
$
%\end{equation}
The gauge transformation is then used on $H_0$ (with ${\bf A}=0$)
and $H_{\rm int}^{\rm mf}$. Upon summing over $\rr$ we keep only
slowly varying terms, {\it i.e.\/} terms where
$\exp[i(\vQ_{\lambda'} \pm \vQ_{\lambda})\cdot\rr]$ vanish. Not
surprisingly, the only surviving terms are either diagonal in
$\lambda$ or have $\vQ_{\lambda'} = \vQ_{\lambda} + \vQ \equiv
\bar{\lambda}$. This means that $\psi_{1\sigma}(\rr)$ and
$\psi_{3\sigma}(\rr)=\psi_{\bar{1}\sigma}(\rr)$ form one subspace,
and $\psi_{2\sigma}(\rr)$ and $\psi_{4\sigma}(\rr) =
\psi_{\bar{2}\sigma}(\rr)$ form the other. Thus it becomes natural
to consider the spinors
%\begin{equation} \label{lambdaspinor}
$
\Psi^{\dagger}_{\lambda}(\rr) = \{
\psi^{\dagger}_{     \lambda \uparrow}(\rr),
\psi^{       }_{\bar{\lambda}\downarrow}(\rr),
\psi^{\dagger}_{\bar{\lambda}\uparrow}(\rr),
\psi^{       }_{     \lambda \downarrow}(\rr)\}.
$
%\end{equation}
The gauge factor
$e^{i\vQ_\lambda\cdot\rr}$ leads to a sign change between the terms
$\psi_{\lambda\sigma}(\rr+\vd_j)$ and
$\psi_{\lambda\sigma}(\rr-\vd_j)$ in $H_0$ and
$\Delta$. The difference terms arising from this becomes derivatives
in the continuum limit. Further care is necessary regarding extra
$\vQ_{\lambda}$-dependent signs. For simplicity we assume ${\bf
m}(\rr) = m(\rr) {\bf e}_z$ and obtain a final Hamiltonian for
the $\Psi_1$ spinor ($\lambda=1$ and $\bar{\lambda}=3$) containing
both $\Delta(\rr)$ and $m(\rr)$:

\end{multicols}
\noindent \lineleft
\begin{equation} \label{HPsi1}
H(\Psi_1) = \Psi_1^{\dagger}(\rr)
\left(
\begin{array}{cccc}
t(i\partial_x+i\partial_y) & \Delta(\rr)^*(i\partial_x-i\partial_y)
& m(\rr) & 0
\\
\Delta(\rr)(i\partial_x-i\partial_y) & -t(i\partial_x+i\partial_y)
& 0 & -m(\rr)
\\
m(\rr) & 0 &
-t(i\partial_x+i\partial_y) & -\Delta(\rr)^*(i\partial_x-i\partial_y)
\\
0 & -m(\rr) &
-\Delta(\rr)(i\partial_x-i\partial_y) & t(i\partial_x+i\partial_y)
\end{array}
\right) \Psi_1(\rr).
\end{equation}
\hfill \lineright
\begin{multicols}{2}
\noindent
With the Ansatz $\Psi_1^{\dagger}(\rr) =
(a_1,a_2,a_3,a_4) e^{-i\vk\cdot\rr}$, and $k_{\pm}= k_x \pm k_y$
for $\vk = (k_x,k_y)$ the eigenvalue problem becomes
\begin{equation} \label{evalprob}
\left| \begin{array}{cccc}
tk_+ - E & \Delta^* k_- & m & 0  \\
\Delta k_- & -tk_+ - E  & 0 & -m \\
m &  0 & -tk_+ - E & -\Delta^* k_- \\
0 & -m & -\Delta k_- & tk_+ - E
\end{array} \right| = 0,
\end{equation}
A pure SC phase has a constant SC order parameter $\Delta$, while
$m=0$, and the spectrum becomes
$E=\pm \sqrt{t^2k_+^2 + |\Delta|^2 k_-^2}$.
The corresponding eigenstates are easily found. For a pure AF phase
$m$ is a constant and $\Delta=0$. The spectrum now becomes
$E=\pm \sqrt{t^2k_+^2 + m^2}$, with associated eigenstates.

We now imagine the plane to be divided into two parts. For $x<0$ the
system is in the SC phase while for $x>0$ it is in the AF
phase. It is now simple to study the scattering problem where a
particle with energy $|E|<m$ in the SC sector is moving towards
the barrier formed by the AF sector. The result is not surprising:
if the particle starts out with a momentum near, say,
$\vQ_{\lambda}$ it is reflected completely by the AF sector (where it
only acquires an exponentially damped probability), and it ends up
with a momentum near either $\vQ_{\lambda}$ or
$\vQ_{\bar{\lambda}}$. The process resembles Andreev
reflection in the quantum number $\lambda$. The conclusion of this
exactly solvable model is clear: low energy particles in the SC sector
can be confined by a surrounding AF sector, or conversely, the AF
sector expels low energy particles.

%-----------------------------------------------------------------------
%\section{Abrikosov vortices in SO(5)}
%\label{sec:vortex_core}
%-----------------------------------------------------------------------

We now proceed to discuss dSC vortices, first briefly mentioning
the case of normal cores followed by our SO(5) model calculation
of vortices with AF cores. In his semiclassical analysis of the
electronic density of states produced by $d$-wave vortices,
Volovik showed \cite{volovik} that only a small part of the
density of states results from quasiparticles localized at the
vortex cores, and that that part is a function of the vortex
density. Hence, in any realistic calculation of quasiparticle
states, the entire vortex lattice must be taken into account. Wang
and MacDonald\cite{wang} made the first self-consistent, numerical
lattice calculation of a tight-binding model for $d$-wave type-II
superconductors using the Bogoliubov-de Gennes equations. In the
following we expand their work to the case of $d$-wave
superconductors with antiferromagnetic cores as described by the
SO(5) model. We take advantage of earlier self-consistent
calculations of isolated vortices in the SO(5)
model,\cite{arovas,bruus} where in the symmetric gauge ${\bf
A}_0(\rr) = \frac{\hbar}{2e}\frac{\alpha_0(r)}{r} {\bf
e}_{\theta}$ the SC and AF order parameters as a function of the
distance to the vortex core in polar coordinates are given by
$\Delta(\rr) = f_0(r) e^{i\theta(r)}$ and $m(\rr) = m_0(r)$,
respectively. Here the functions $f_0(r)$, $\alpha_0(r)$ and
$m_0(r)$ are only known numerically. Note that ${\bf m}(\rr)$
points in a constant direction, so only the size $m(\rr)$ is
given. For a lattice of non-overlapping vortices, {\it i.e.\/}
vortices further than a few times the London length apart, the
self-consistent solutions for vector potential, the SC and the AF
order parameter are expressed in term of the vortex centers,
$\vbr_j$, as
\begin{eqnarray}
\label{SCorder}
\Delta(\rr) &=& f(\rr) e^{i\theta(\rr)} =
\prod_j f_0(\rr-\vbr_j) e^{\sum_k i{\rm Arg}(\rr-\vbr_k)}\\
\label{AForder}
m(\rr) &=& \sum_j m_0(\rr-\vbr_j)\\
\label{Vecpot}
{\bf A}(\rr) &=& \sum_j {\bf A}_0(\rr-\vbr_j),
\end{eqnarray}
where ${\rm Arg}(\rr-\vbr_k)$ is the polar angle between $\rr$ and
$\vbr_k$. In a lattice model a particularly simple way to construct
the magnetic unit cell is the following. For each area penetrated by
one flux quantum $h/e$ a Dirac anti-vortex string carrying a flux
$-h/e$ is added.\cite{AddDirac} The strings will have no physical
consequences at all when placed between lattice sites. However, they
allow for the construction of a vector potential periodic in the
magnetic unit cell, through which the magnetic flux is zero.

We now construct a square Abrikosov lattice with a Dirac anti-vortex
added to the center of every second vortex. Since each vortex carry
half a flux quantum, the smallest magnetic unit cell possible
consists of two vortices. However, due to better convergence
properties in obtaining the periodic vector potential and a periodic
representation of the SC order parameter (especially its phase
$\theta(\rr)$), we choose to double the magnetic unit cell.
Our unit cell contains two ordinary vortices on one diagonal and
two vortices penetrated by Dirac anti-vortices on the other. Periodic
forms of ${\bf A}(\rr)$, $\theta(\rr)$, and $m(\rr)$ are then easily
found by adding up contributions from a large number of unit cells
(typically 64) surrounding the one we are studying. From this we
obtain a mean field lattice Hamiltonian $H = H_0 + H^{\rm mf}_{\rm
int}$ given by Eqs.~(\ref{H0}) and~(\ref{Hmfint}), which is periodic
in our unit cell. Based on the Bogoliubov
transformation for operators within our unit cell
\begin{equation} \label{bogoliubov}
(\gamma^{\alpha}_{\sigma})^{\dagger} = \sum_{\rr} \left\{
u^{\alpha}(\rr) c^{\dagger}_{\sigma}(\rr) + \sigma
v^{\alpha}(\rr) c_{\bar{\sigma}}(\rr)
\right\},
\end{equation}
where $\sigma = \pm1$ is the spin index and $\bar{\sigma} = -\sigma$,
the equation of motion for the $\gamma^{\alpha}_{\sigma}$-operators
using the periodic Hamiltonian $H$ leads to the Bogoliubov-de Gennes
equation for the eigenenergies and eigenstates of the fermionic
quasiparticles:
\begin{equation} \label{bdg}
\left( \begin{array}{cc}
T + \sigma M & D \\
D^* & -T^* + \sigma M
\end{array} \right)
\left( \begin{array}{c}
{\bf u}^{\alpha}\\
{\bf v}^{\alpha}
\end{array} \right)
= E^{\alpha}
\left( \begin{array}{c}
{\bf u}^{\alpha}\\
{\bf v}^{\alpha}
\end{array} \right).
\end{equation}
Here $E^{\alpha}$ is the quasiparticle energy,
${\bf u}^{\alpha}$ and ${\bf v}^{\alpha}$ are
vectors containing the values of $u^{\alpha}(\rr)$
and $v^{\alpha}(\rr)$ on each lattice site in our unit cell,
while the block matrices $T$, $D$, and $M$ are given by
\begin{eqnarray}
\label{Tmatrix}
(T)_{\rr\rr'} & = &
-te^{-i\frac{e}{\hbar}\int_{\rr}^{\rr'}\!
{\bf A}\cdot d{\bf l}} \sum_{j=1}^{4} \delta_{\rr',\rr+\delta_j} \\
\label{Dmatrix}
(D)_{\rr\rr'} & = &
\sum_{j=1}^{4} \phij  [D(\rr')+D(\rr)]
\delta_{\rr',\rr+\delta_j}\\
(M)_{\rr\rr'} & = &
\sum_{j,j'=1}^{4} \phij \phijp M(\rr' - \vd_j)
\delta_{\rr',\rr+\delta_j+\delta_{j'}}\nonumber\\
\label{Mmatrix}
&& -M(\rr')\delta_{\rr',\rr},
\end{eqnarray}
with $D(\rr) = \frac{1}{2} V \langle \Delta(\rr) \rangle$ and
$M(\rr) = \frac{1}{2} e^{i\vQ\cdot\rr}V \langle m(\rr) \rangle$.

\begin{figure}
\centerline{\epsfxsize=\linewidth\epsfbox{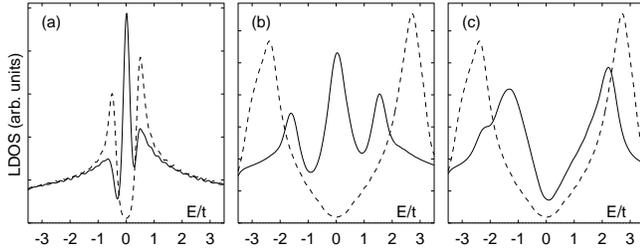}}
\narrowtext
\caption{ \label{fig:LDOS}
LDOS in the vortex core (full line) and in the bulk SC (dashed line)
for (a) BCS $s$-wave SC, (b) BCS $d$-wave SC, and (c) SO(5)
SC with an AF vortex core.}
\end{figure}

In the numerical calculation we use a $N\!\times\!N$ lattice with
$N=44$. The origin is put in the center and the four vortices in
the center of each of the quadrants. The periodicity is ensured by
having $H(\rr+N\vd_1) = H(\rr+N\vd_2) = H(\rr)$. The Bogoliubov-de
Gennes equation, Eq.~(\ref{bdg}) becomes a $2N^2\times2N^2$
eigenvalue problem yielding for a given value of the spin variable
$\sigma$ the spectrum $E^{\alpha}$ and the Bogoliubov coefficients
${\bf u}^{\alpha}$ and ${\bf v}^{\alpha}$. To compare our
calculations with the experimental STM measurements on
vortices\cite{hess,fischer,renner} and with the existing
calculations\cite{wang} on ordinary sSC and dSC vortices we
compute the temperature dependent local density of states (LDOS)
according to the standard minimal model\cite{shore,gygi,wang}

\begin{eqnarray}
N(\rr,E) &=& \sum_{\alpha} [
|u^{\alpha}(\rr)|^2 \{-f'(E^{\alpha} - E)\}
\nonumber \\ \label{LDOS} &&+
|v^{\alpha}(\rr)|^2 \{-f'(E^{\alpha} + E)\}],
\end{eqnarray}
where $f(\varepsilon)=[\exp(\varepsilon/k_BT)+1]^{-1}$, and where we
have neglected the dispersion in the magnetic Brillouin zone.
The calculation yields the LDOS shown in
Fig.~\ref{fig:LDOS}. In all cases $V = 0.8t$, $k_{\rm B}T =
0.1t$  and  $\mu = -0.6t$, which due to the band structure leads to
an asymmetric LDOS.
First, to check our calculations, we change the model
from SO(5) to ordinary sSC and dSC. The latter is produced by
setting $M_{\rr,\rr'} = 0$ in Eq.~(\ref{Mmatrix}), and the former by
furthermore setting $(D)_{\rr,\rr'} =
\frac{1}{2}V\langle\Delta(\rr)\rangle \delta_{\rr,\rr'}$ in
Eq.~(\ref{Dmatrix}). As shown in
Fig.~\ref{fig:LDOS}a and~\ref{fig:LDOS}b we confirm qualitatively the
main conclusions of Refs.~\onlinecite{wang,franz,yasui}. In the bulk
of the sSC phase a full gap is observed, while a mid-gap peak (which
splits at $T=0$) develops in the center of a sSC vortex. In the bulk
of the dSC phase a steady rise of the LDOS is seen around the mid-gap
position, while a mid-gap peak develops in the center of a dSC
vortex. Our model calculation captures mainly generic features and can
therefore not be used in the ongoing debate of the detailed form of
the LDOS in the dSC vortex core.\cite{wang,franz,yasui} However, this
issue is not important for our main observation in the SO(5) case:
instead of a mid-gap peak the LDOS is dramatically suppressed in the
AF vortex core resembling bulk behavior as shown in
Fig.~\ref{fig:LDOS}c. This confirms the conclusion of the dSC/AF
interface in the SO(5) model studied in the first part of this
paper. The AF phase effectively suppresses any fermionic low energy
states.

%-----------------------------------------------------------------------
%\section{Concluding remarks}
%\label{sec:conclusion}
%-----------------------------------------------------------------------
We thus reach our main conclusion. The experimentally observed lack of
electronic quasiparticle states in the center of Abrikosov vortices in
BSCCO\cite{renner} as opposed to the measurements of a normal metallic
core of vortices in YBCO\cite{fischer} finds a natural explanation in
the framework of the SO(5) model. As already pointed out by Arovas et
al.\cite{arovas} the nature of the SO(5) vortex cores are governed by
the parameters ({\it e.g.\/} doping level and coupling strengths) of
the given high-$T_c$ material. The cores can either become metallic,
{\it i.e.\/} a pure dSC behavior, or insulating, {\it i.e.\/} a mixed
dSC/AF behavior. At the present stage of the SO(5) theory it is
difficult to predict which materials will in fact develop AF vortex
cores. For example, as is studied in the striped phase\cite{berlinsky},
the insulating vortex cores are negatively charged, since they must be
at half filling, in contrast to the hole doped bulk material maintained at
lower filling. Such a charging energy must be taken into account in a
detailed calculation of the energy gained by forming an AF vortex core.
Our calculation of the generic features in the fermionic sector of the
SO(5) model shows that the measured LDOS can be explained if one simply
assumes that YBCO with its metallic vortex cores is a pure dSC SO(5)
superconductor, while BSCCO is a dSC/AF SO(5) superconductor. We
obtained our results by studying both the analytically solvable model
of a perfect SC/AF interface and by exact numerical diagonalization of
an Abrikosov lattice model. Clearly, further theoretical
insight in the dual dSC/AF nature of the high-$T_c$ compounds can be
obtained from studies of the striped phases, where alternating stripes
of SC phases and AF phases occur\cite{berlinsky}.

%-----------------------------------------------------------------------
%\section{Acknowledgements}
%\label{sec:acknowledgements}
%-----------------------------------------------------------------------
This work was supported by the Danish Natural Science Research
Council: access to the Cray 92 at Uni$\bullet$C through Grant
No.\ 9602481, and H.B.\ through Ole R\o mer Grant No.\ 9600548.

\end{multicols}

\begin{references}
\bibitem{AddDirac} Mathematically, the Dirac string is added to a
  point $\rr_0$ by the gauge transformation $\chi(\rr) = -(\hbar/e){\rm
  Arg}(\rr-\rr_0)$.
\end{references}
\end{document}